\documentclass[10pt,
aps,
pra,
amsmath,amssymb,
twocolumn,
superscriptaddress,
amsmath,amssymb,
]{revtex4-2}



\usepackage{nameref}
\usepackage{amsmath}
\usepackage{titlesec}
\usepackage{graphicx}
\usepackage{dcolumn}
\usepackage{bm}
\usepackage{rotating}
\usepackage{mathrsfs}
\usepackage{mathdots}
\usepackage[dvipsnames]{xcolor}
\usepackage{lineno}
\usepackage[colorlinks,
 linkcolor = black,
 citecolor = red,
 urlcolor = black,
 anchorcolor = black,
 filecolor = black,
 ]{hyperref}
\begin{document}
\titlespacing{\section}
 {0pt}
 {10pt}
 {0pt}
\titleformat*{\section}{\Large\bfseries\raggedright}
\titlespacing{\subsection}
 {0pt}
 {15pt}
 {5pt}
\titleformat{\subsection}[block] 
 {\bfseries}
 {}
 {0pt}
 {}

\title{Pump-Threshold-Free Frequency Comb via Cavity Floquet Engineering}

\author{Sihan Wang}
\affiliation{Beijing Key Laboratory of Fault-Tolerant Quantum Computing, Beijing Academy of Quantum Information Sciences, Beijing 100193, China}
\affiliation{Hefei National Research Center for Physics Sciences at Microscale, University of Science and Technology of China, Hefei 230026, China}
\affiliation{Laboratory of Spin Magnetic Resonance, University of Science and Technology of China, Hefei 230026, China}

\author{Cheng Wang}
\affiliation{Department of Applied Physics, Aalto University, FI-00076 Aalto, Finland}

\author{Matthijs H.~J.~de Jong}
\affiliation{Department of Applied Physics, Aalto University, FI-00076 Aalto, Finland}

\author{Laure Mercier de L\'epinay}
\affiliation{Department of Applied Physics, Aalto University, FI-00076 Aalto, Finland}

\author{Jingwei Zhou}
\email{zhoujw@ustc.edu.cn}
\affiliation{Hefei National Research Center for Physics Sciences at Microscale, University of Science and Technology of China, Hefei 230026, China}
\affiliation{Laboratory of Spin Magnetic Resonance, University of Science and Technology of China, Hefei 230026, China}

\author{Mika A. Sillanp\"{a}\"{a}}
\affiliation{Department of Applied Physics, Aalto University, FI-00076 Aalto, Finland}

\author{Yulong Liu}
\email{liuyl@baqis.ac.cn}
\affiliation{Beijing Key Laboratory of Fault-Tolerant Quantum Computing, Beijing Academy of Quantum Information Sciences, Beijing 100193, China}

\begin{abstract}
Frequency combs have revolutionized communication, metrology, and spectroscopy. Considerable efforts have been devoted to developing integrated combs, primarily leveraging Pockels or Kerr nonlinearities. Here, we demonstrate an alternative frequency comb generated via cavity Floquet engineering. By periodically modulating the cavity resonance frequency through a driven mechanical oscillator, a Floquet cavity with multiple equally spaced frequency components is created. These sidebands exhibit nearest-neighbor coupling and are phase-locked to the external modulation drive. A pump tone interacts with the pre-modulated cavity to generate the output frequency comb, which we implement in an on-chip microwave cavity optomechanical system. This approach operates independently of a pumping threshold and is insensitive to pump detuning. Consequently, it enables comb generation under far-sideband pumping with nanowatt-scale total power consumption, providing an ultra-low-power platform for integrated frequency comb synthesis.
\end{abstract}
\maketitle

\section*{Introduction}

Frequency combs, composed of discrete, equally spaced frequencies~\cite{diddams2020Optical, chang2022Integrated}, have contributed to advancements in optical communication~\cite{marin-palomo2017Microresonatorbased, jorgensen2022Petabitpersecond}, precision metrology~\cite{udem2002Optical, liang2025Modulated}, spectroscopy~\cite{picque2019Frequency, han2024Dualcomb}, and atomic clocks~\cite{takamoto2005Optical, roslund2024Optical, wu2025Vernier}. The compact on-chip frequency combs (microcombs) mainly consist of Kerr combs~\cite{delhaye2007Optical, kippenberg2011MicroresonatorBased, kippenberg2018Dissipative} and electro-optic (EO) combs~\cite{zhang2019Broadband, hu2022Highefficiency, yu2022Integrated, stokowski2024Integrated, zhang2025Ultrabroadband}. Kerr combs relying on $\chi^{3}$ nonlinearity have been successfully demonstrated in high-$Q$ microresonators based on materials such as silica~\cite{delhaye2007Optical}, silicon nitride~\cite{shen2020Integrated}, silicon carbide~\cite{guidry2022Quantum}, diamond~\cite{hausmann2014Diamond}, and lithium niobate (LN)~\cite{wang2019Monolithic}. Integrated EO combs utilize the Pockels effect, which has also recently been realized in thin-film LN devices~\cite{zhang2019Broadband, hu2022Highefficiency,yu2022Integrated} and lithium tantalate~\cite{zhang2025Ultrabroadband}. Despite these remarkable advances, EO and Kerr combs are facing limitations of their generation efficiency~\cite{zhang2019Broadband,yi2016Theory, xue2019Superefficient} or tuning precision~\cite{herr2014Temporal, shen2020Integrated}. Therefore, emerging paradigms for frequency comb generation are highly desired~\cite{flower2024Observation, piccardo2020Frequency, chou2020Frequencycomb}, as they offer promising avenues to overcome the aforementioned limitations in applications such as astrocombs~\cite{obrzud2019Microphotonic, cheng2024Continuous} and molecular fingerprinting~\cite{picque2019Frequency, schliesser2012Midinfrared}. The frequency combs based on engineered nonlinearity could also offer a scalable and low-cost manner for emerging fields such as integrated photonic data links~\cite{shu2022Microcombdriven}, quantum networks~\cite{roslund2014Wavelengthmultiplexed}, and multiphoton entanglement~\cite{xie2015Harnessing, reimer2016Generation, jia2025Continuousvariable}. Notably, Floquet engineering~\cite{bukov2015Universal}, which involves the periodic time-dependent modulation, has the capability to control and generate multi-mode sidebands by tuning energy spectra~\cite{li2013Motional, else2016Floquet, mercierdelepinay2020Nonreciprocal, lukin2020Spectrally, ito2023Buildup, zhou2021Rapid, zhou2024Cavity}, as well as to induce interactions between modes~\cite{dutt2019Experimental, clark2019Interacting}. These equally spaced sidebands demonstrate that Floquet engineering is a promising, albeit underexplored, paradigm for frequency comb generation~\cite{qiu2025Floquet,he2025coherent}, without necessitating Kerr nonlinearity or Pockels effect.
\par
In this work, we demonstrate a frequency comb generated via cavity Floquet engineering, termed the Floquet cavity frequency comb. Specifically, by periodically modulating the resonance frequency of a cavity, we realize a Floquet cavity with multiple equally spaced sidebands. These sidebands are coupled via nearest-neighbor interactions and can be further phase-locked to the external periodic modulation. Then, a single-frequency pump tone is added, interacting with this pre-modulated multi-mode structure to generate the output frequency comb. Additionally, the absolute positions of the comb teeth are referenced to the frequency of the pump tone, with the intensity approximately symmetrically distributed around the cavity's intrinsic frequency. Furthermore, in contrast to Kerr or EO combs based on microresonators, which require precise frequency tuning to match the free spectral range, our approach is flexibly tunable, with the spacing and number of comb teeth determined by the modulation source. The frequency components of the Floquet cavity are internally predefined, enabling the frequency comb to operate without the need to surpass a pumping threshold.
\par
\begin{figure*}
\centering
\includegraphics[width=1\linewidth]{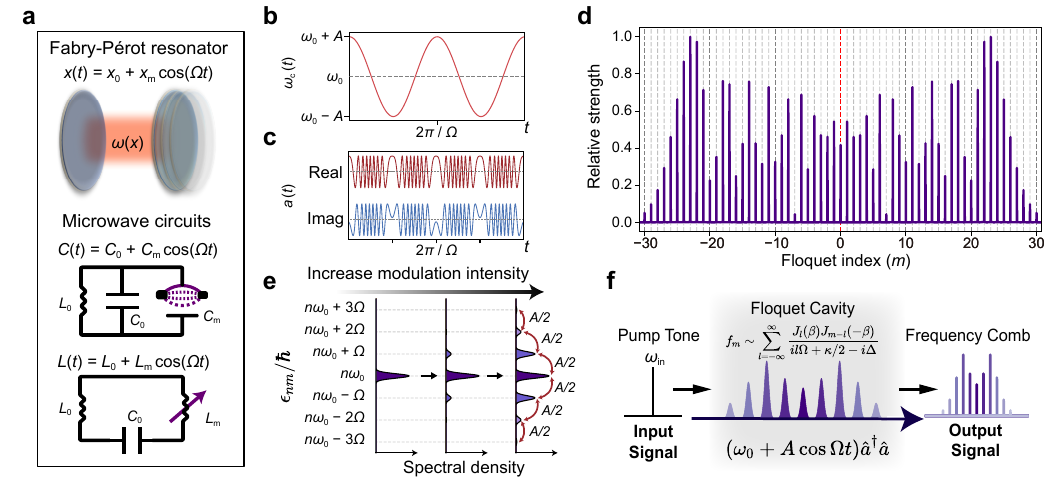}
\caption{\label{fig1}
\textbf{Floquet cavity and frequency comb.}
\textbf{a} Examples of cavities with periodically modulated resonance frequencies.
\textbf{b} Model of the Floquet cavity with the modulation frequency $\Omega$ and strength $A$.
\textbf{c} Time evolution of the intracavity field amplitude in the Floquet cavity obtained via numerical simulation. For demonstration purposes, cavity dissipation is neglected and $ A \approx 25 \Omega $.
\textbf{d} The relative strength in the frequency domain of the Floquet cavity modes $ \omega_0 + m \Omega $ (where $m$ is the index of the Floquet states). The red dotted line marks the intrinsic frequency $\omega_0$.
\textbf{e} Schematics of the spectral density of Floquet cavity quasi-energy with increasing modulation strength $A$ (where $n$ is the Fock basis). \textbf{f} The schematic diagram of the generation of Floquet cavity frequency comb. By injecting the pump tone into the pre-modulated cavity, photons are emitted across multiple modes, resulting in a frequency comb.
}
\end{figure*}
Based on our theoretical framework, such a frequency comb can be realized across a wide range of physical systems. Here, we demonstrate its implementation for modulating the resonance frequency of a superconducting microwave cavity via a mechanically compliant capacitor. To achieve sinusoidal modulation, we additionally employ an auxiliary cavity to establish a conventional optomechanical coupling, driving the mechanical oscillator from thermal state into self-induced oscillation~\cite{aspelmeyer2014Cavity}. In contrast to Kerr optomechanical combs~\cite{aldana2013Equivalence, gong2009Effective, miri2018Optomechanical}, which rely on the nonlinear coupling between phonons and photons~\cite{hu2021Generation, shin2022OnChip, wu2022Hybridized}, our approach achieves an extended tuning range for the pump signal, including frequencies that are substantially red-detuned from the cavity's natural resonance. Furthermore, this approach reduces the pump power by a factor of approximately $10^6$, which is limited primarily by the detection efficiency. Cavity Floquet engineering enables efficient frequency comb generation even when the pump is far detuned from the cavity’s intrinsic frequency, with the total power reaching the nanowatt level. This thus provides an ultra-low-power platform for frequency comb generation.
\par
\section*{Results}
\subsection*{Floquet cavity frequency comb}

We introduce a general Floquet cavity, defined as a cavity whose resonance frequency $\omega_\mathrm{c}$ is periodically modulated in time. It can be mapped to counterparts in well-established physical systems, including Fabry--Pérot resonators of a movable mirror or microwave circuits with adjustable capacitance or inductance, as illustrated in Fig.~\ref{fig1}\textbf{a}. The resonance modes of these Floquet cavities transform from a single-frequency state to a superposition of multiple frequencies (see Supplementary Note~I.A). The Hamiltonian of the Floquet cavity is given by $H_{\mathrm{Floquet}} = \mathrm{\hbar} \left(\omega_\mathrm{0} + A \cos(\Omega t+\phi_0)\right)\hat{a}^\dagger \hat{a}$, where $\omega_\mathrm{0}$ is the intrinsic frequency, $A$ is the modulation strength, $\Omega$ is the modulation frequency, and we drop an irrelevant phase $\phi_0$, as illustrated in Fig.~\ref{fig1}\textbf{b}.
\par
In terms of Floquet theory, such periodic modulation of the cavity frequency effectively creates ``Floquet sidebands'' around the cavity's intrinsic frequency. The quasi-energy of this Floquet cavity satisfies $\epsilon_{nm} /\hbar= n\omega_\mathrm{0} + m\Omega$, with the eigen-frequencies $ \omega_\mathrm{0}, \omega_\mathrm{0} \pm \Omega, \omega_\mathrm{0} \pm 2\Omega, \dots $ (see Methods), where integer $n$ is the Fock basis of bare cavity and $m$ is the Floquet index. We present a numerical example illustrating the time evolution of the intracavity field in the Floquet cavity in Fig.~\ref{fig1}\textbf{c}. A set of equally spaced components corresponding to the intracavity field in the Fourier domain is shown in Fig.~\ref{fig1}\textbf{d}. The relative strength of the Floquet modes is determined by $ J_{m} (\beta)$, the Bessel function of the first kind with order $ m $ (see Supplementary Note~I.B and I.C). Here we define the dimensionless modulation coefficient $ \beta = \frac{A}{\Omega} $. With a sufficiently large modulation strength, a substantial number of detectable Floquet modes can be produced.
\par
Remarkably, since these modes are induced by Floquet engineering, they are not independent (as in modes in spatially separate cavities) but are correlated. The modulation strength $ A/2 $ determines the adjacent coupling strength between the different Floquet modes, thus forming what is referred to as a giant-mode cavity. A stronger modulation strength clearly enhances the generation of additional sidebands with appreciable intensity, as shown in Fig.~\ref{fig1}\textbf{e}. Furthermore, this coupling enables energy input into the cavity to be transferred to different sidebands. It is notable that these sidebands inherently exhibit coherence, with their phases locked to the external modulation. Consequently, any input pump tone, even in the few-photon regime, is emitted across multiple sideband modes, forming a frequency comb with spacing $\Omega $, as illustrated in Fig.~\ref{fig1}\textbf{f}.
\par
Considering an input pump tone at the frequency $ \omega_{\mathrm{in}} $, from the equation of motion, we obtain that the complex amplitude in the Floquet cavity satisfies (Supplementary Note~II.A):
\begin{equation}\label{eqn-1}
\begin{aligned}
a(t)= &\sqrt{\kappa_{\mathrm{e}}} a_{\mathrm{in}} e^{-i \Delta t} \\
 &\times \sum_{m=-\infty}^{\infty} \sum_{l=-\infty}^{\infty} \frac{ J_{l}(\beta) J_{m-l}(-\beta) }{i l\Omega + \kappa/2 - i\Delta}e^{im\Omega t},
\end{aligned}
\end{equation}
where $ \Delta = \omega_{\text{in}} - \omega_0 $ arises from the rotation of the frame, $ \kappa $ represents the total dissipation of the cavity, $ \kappa_\mathrm{e} $ denotes the external dissipation due to the coupling between the cavity and the external input/output, and $ a_\mathrm{in} $ represents the externally injected pump field. For a coherent modulation source and pump tone, the output field consequently forms a set of isolated and coherent frequency-comb lines in the frequency domain. Noise from the pump and modulation sources causes linewidth broadening of the comb teeth, with details provided in Supplementary Notes~II.B and II.C.
\par
For a Floquet cavity frequency comb, the modulation coefficient $ \beta $ is predefined and decoupled from $ a_\mathrm{in} $, enabling frequency comb operation without a power threshold for the pumping tone, as indicated by Equation~\eqref{eqn-1}. In other terms, the pump tone does not induce nonlinear effects but instead serves as a probe for a system with a prescribed time-varying response function. Furthermore, we found that the pump tone exhibits flexible tuning capabilities within the approximate range $ -2\beta \Omega \lesssim \Delta \lesssim 2\beta \Omega $, as detailed in the Supplementary Note~II.A. In contrast, in Kerr optomechanical combs, the generation of the frequency comb by optomechanical non-linearity depends on the pump tone itself, meaning $ \beta $ is influenced by both $a_\mathrm{in}$ and $\Delta$. Consequently, $ a_\mathrm{in} $ must reach a sufficient intensity to overcome an initiation threshold. Additionally, $ \Delta $ needs to be constrained within a range that can effectively excite the oscillator (For $ \kappa < \Omega $, $ \Delta \approx \Omega $) in Kerr optomechanical combs. In terms of the comb spectral profile, EO and Kerr combs are typically symmetric about the pump frequency, whereas the Floquet cavity frequency comb is centered at the intrinsic cavity resonance ($\omega_{0}$), with a spectral envelope that is largely independent of the pump detuning.

\subsection*{The Device}

\begin{figure*}
\centering
\includegraphics[width=1\linewidth]{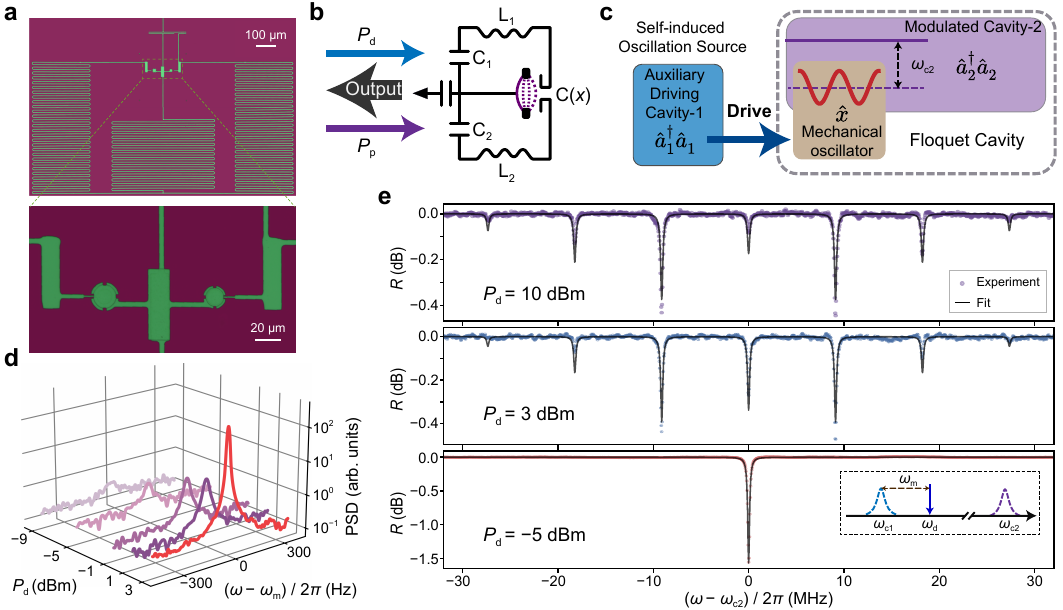}
\caption{\label{fig2}
\textbf{Construct the Floquet cavity in optomechanical systems.}
\textbf{a} Micrographs of the device, showing the entire chip and zoomed-in drums. We select the larger drum (left one) for the demonstration, while the other one is ignored.
\textbf{b} Equivalent circuit model: straight arrows indicate the corresponding applied signals (blue for cavity-1, purple for cavity-2).
\textbf{c} Cavity-1 serves as the auxiliary driving cavity to drive the mechanical oscillator into self-induced oscillation. Cavity-2 is modulated by the oscillator to construct the Floquet giant-mode cavity.
\textbf{d} Mechanical oscillator spectrum transitioning from thermal state to self-induced oscillation regime; the red curve marks the spectrum at the critical drive power $ P_\mathrm{dc} $.
\textbf{e} The reflection spectrum of the cavity-2 measured by S21 as a function of increasing drive power. Colored dots: measurements; black lines: fits based on the reflection coefficient $R$. $ X_\mathrm{0} $ is incorporated into $ \omega_\mathrm{c2} $. The inset marks the driving frequency schematics.
}
\end{figure*}
We demonstrate a Floquet cavity frequency comb in a superconducting microwave circuit, as illustrated in Fig.~\ref{fig2}\textbf{a}. The resonance frequency of the cavity is modulated by a displacement-dependent capacitor. The sample used in this study is similar to that reported in our previous work~\cite{mercierdelepinay2021Quantum, wang2024Groundstate} (detailed fabrication in Supplementary Note~III.A). Our device consists of two aluminum drumhead mechanical oscillators~\cite{teufel2011Circuit, teufel2011Sideband} optomechanically coupled to two microwave cavities, as schematically shown in Fig.~\ref{fig2}\textbf{b}. Here, we ignore one of the mechanical oscillators. The Hamiltonian of such a system can be expressed as
\begin{equation}\label{eqn-2} 
 H_\mathrm{sys}=\sum_{j=1,2}\hbar ( \omega_{\mathrm{c},j} + \frac{g_{\mathrm{0},j}}{x_\mathrm{zpf}}\hat{x})\hat{a}_j^\dagger \hat{a}^{\mathstrut}_j + \hbar \omega_\mathrm{m}\hat{b}^\dagger \hat{b},
\end{equation}
where $ \omega_{\mathrm{c},j} $ represents the $ j $-th cavity's intrinsic frequency. $ \omega_\mathrm{m} $ represents the mechanical resonant frequency, and $ a_j $ ($ a_j^\dagger $) and $ b $ ($ b^\dagger $) represent the annihilation (creation) operators of the cavities and mechanical oscillator, respectively. $\hat{x}$ is the displacement of the mechanical oscillator, and $ x_\mathrm{zpf} $ represents the zero-point fluctuation of the mechanical oscillator’s displacement. Besides, the coefficient $ g_{\mathrm{0},j} $ represents the single-photon coupling strength between the $ j $-th cavity and mechanical oscillator.
\par
Our entire device operates in a dilution refrigerator at a temperature of 10 mK (detailed setup in Supplementary Note~III.B). The first cavity's (cavity-1) intrinsic frequency is $\omega_{\mathrm{c}1} =2\pi\times 4.91~\text{GHz}$, with external decay rate $\kappa_{\mathrm{e}1} = 2\pi \times 49.6~\text{kHz}$ and intrinsic decay rate $\kappa_{\mathrm{i}1} = 2\pi \times 298.9~\text{kHz}$. For the other microwave cavity (cavity-2), the intrinsic frequency is $\omega_{\mathrm{c}2} = 2\pi \times 6.47~\text{GHz}$, accompanied by $\kappa_{\mathrm{e}2} = 2\pi \times 21.7~\text{kHz}$ and $\kappa_{\mathrm{i}2} = 2\pi \times 245.7~\text{kHz}$. The oscillator is characterized to have frequency of $\omega_\mathrm{m} = 2\pi \times 9.1~\text{MHz}$ and the damping rate $\gamma_\mathrm{m} = 2\pi \times 124~\text{Hz}$. Additionally, from the measurements based on the frequency modulation technique, the single-photon coupling strengths are measured to be $g_{0,1} = 2\pi \times 79~\text{Hz}$ and $g_{0,2} = 2\pi \times 46~\text{Hz}$. The detailed characterization method is described in Supplementary Note~III.C.

\subsection*{Construct the Floquet cavity }

We choose the cavity-2 and the mechanical oscillator together to form the Floquet cavity system with modulation frequency $\omega_\mathrm{m}$, as schematically shown in Fig.~\ref{fig2}\textbf{c}. It is worth noting that, in the absence of excitation, the mechanical oscillator is in a thermal state in equilibrium, with an extremely small time-averaged amplitude and random phase. To achieve Floquet modulation of cavity-2 in a sinusoidal periodic manner, we designate cavity-1 as an auxiliary driving cavity to drive the mechanical oscillator into a self-induced oscillation (Supplementary Note~IV) through optomechanical parametric excitation~\cite{aspelmeyer2014Cavity, marquardt2006Dynamical, ludwig2008Optomechanical, metzger2008SelfInduced}.
\par
We apply a blue-sideband drive at the frequency $\omega_{\mathrm{d}} = \omega_{\mathrm{c1}} + \omega_\mathrm{m}$ with drive power $P_\mathrm{d}$. As the drive power increases, the mechanical oscillator’s vibrations are gradually amplified and enter self-induced oscillation at a critical power $(P_\mathrm{dc})$ of $ 3~\mathrm{dBm} $, as shown in Fig.~\ref{fig2}\textbf{d}. Note that all reported power values correspond to the output power of the signal generator. The mechanical displacement of the self-induced oscillation can be described as a coherent oscillatory motion $x(t) = X_\mathrm{m} \cos(\omega_\mathrm{m} t + \phi_\mathrm{m}) + X_0$. Here, $X_\mathrm{m}$ denotes a stable oscillation amplitude and $X_0$ is the equilibrium offset incorporated into the effective cavity resonance $\omega_\mathrm{c2}$ (see Methods). Correspondingly, the modulation strength $A = \frac{g_{02} X_\mathrm{m}}{x_\mathrm{zpf}}$ is governed by the external drive power $P_\mathrm{d}$. Furthermore, the oscillation phase exhibits coherence with the blue-sideband drive, arising from the underlying optomechanical phonon lasing mechanism~\cite{marquardt2006Dynamical,ludwig2008Optomechanical,metzger2008SelfInduced,krause2015Nonlinear}, which sustains stable periodic motion.
\par
We characterize the Floquet sidebands of cavity-2 by using a vector network analyzer to extract the reflection coefficient $ S_\mathrm{21} = R = 1 - \sqrt{\kappa_\mathrm{e}} \frac{\langle a(t) \rangle}{\langle a_\mathrm{in}(t) \rangle} $. Combining with Equation~\eqref{eqn-1}, the depth of the different sidebands is also determined by the Bessel function, as shown in Fig.~\ref{fig2}\textbf{e}. These sidebands are symmetrically distributed with a spacing corresponding to the oscillator frequency $ \omega_{\mathrm{m}} $. We studied the effect of drive power, and the results are shown in Fig.~\ref{fig2}\textbf{e} for $ P_\mathrm{d} $ at $-5$~dBm, 3~dBm, and 10~dBm. Before the critical power point $ P_\mathrm{dc} $, the mechanical oscillator stays in a thermal state, and thus only the intrinsic resonance frequency of cavity-2 can be detected. When $ P_\mathrm{d} > P_\mathrm{dc} $, the self-induced mechanical oscillation, with stable amplitude and phase, modulates cavity-2 such that it can be effectively described as the Floquet cavity. Increasing the modulation strength enhances the ability of photons to transition to higher-order Floquet energy levels. Consequently, the intensity of the intrinsic frequency signal diminishes, while higher-order sidebands become more pronounced. The result from the free parameter fitting indicates that, under a 10 dBm drive, $ \beta$ is approximately 1.92. Given the other device parameters, the corresponding mechanical oscillator amplitude $X_\textrm{m}$ is calculated as 1.2~nm.
\begin{figure*}
\centering
\includegraphics[width=1\linewidth]{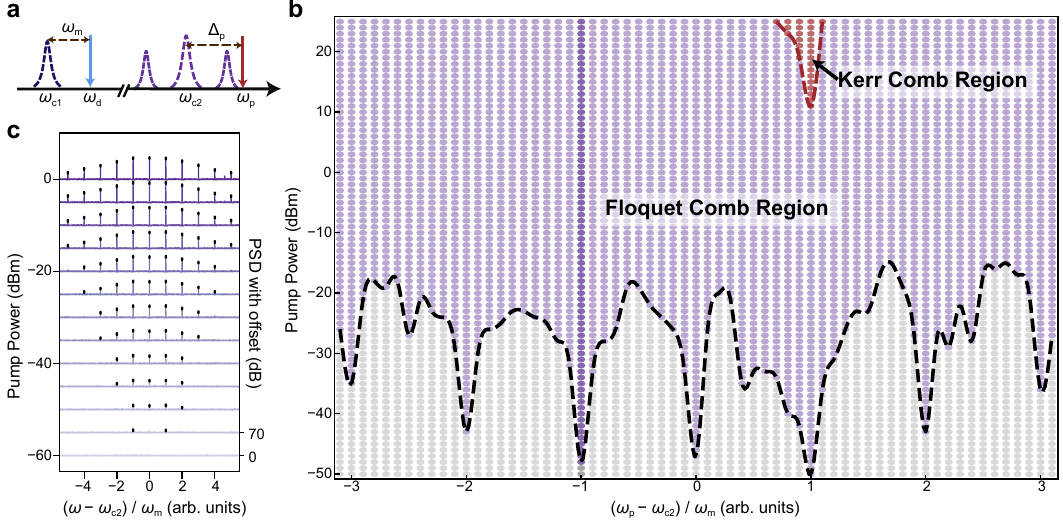}
\caption{\label{fig3}
\textbf{Emergence and evolution of Floquet cavity frequency comb.}
\textbf{a} Illustration of the setup. The blue arrow represents a fixed drive signal at $ \Delta_\mathrm{d}=\omega_\mathrm{m}$ to excite the mechanical oscillator. The red arrow in the Floquet cavity (cavity-2) represents the pump signal responsible for the generation of the frequency comb.
\textbf{b} Emergence and evolution of frequency combs. Spectral responses are sorted into three types and are marked in different colors as dots. We consider the frequency comb to be effective when the number of comb teeth $\ge 4$. The gray points represent the absence of the comb. The Floquet comb manifests in the regions marked with purple dots. In the red-dotted region, cavity-2 itself produces a Kerr frequency comb. The black and red dashed lines are the dividing lines between the different regions.
\textbf{c} Several examples of frequency comb corresponding to the dark points highlighted in \textbf{b} are shown, with each comb tooth marked by a line and a dot.
}
\end{figure*}
\subsection*{Emergence and evolution of Floquet comb}

To achieve a stable self-induced oscillation, we applied a fixed drive power $ P_\mathrm{d} = 10 $ dBm. As shown in Fig.~\ref{fig3}\textbf{a}, we send a pump tone to cavity-2 and characterize the emergence and evolution of the frequency comb in the cavity response signal with different detuning $\Delta_\mathrm{p}=\omega_\mathrm{p} - \omega_\mathrm{c2} $ and power $P_{\mathrm{p}}$ (see Fig.~\ref{fig3}\textbf{b}). The local minima occurring at the sideband frequencies $\Delta_\mathrm{p} = m \omega_\mathrm{m}$ are since the pump frequency matches one of the Floquet sidebands, thereby allowing efficient photon injection. We demonstrate the flexible tuning capability of the Floquet frequency comb within the range of $ -3\omega_\mathrm{m} \leq \Delta_\mathrm{p} \leq 3\omega_\mathrm{m} $. Notably, frequency comb can also emerge in the red sideband region of the cavity-2 ($\Delta_\mathrm{p} < 0$), which is completely absent in Kerr optomechanical combs (for $ \kappa < \Omega $)~\cite{miri2018Optomechanical, hu2021Generation, shin2022OnChip, wu2022Hybridized, krause2015Nonlinear}. In theory, the pump power could be much lower, but it is constrained by the limitations of the detection accuracy in our experimental setup. 
Fig.~\ref{fig3}\textbf{c} shows an example of the frequency comb at a pump detuning of $\Delta_\mathrm{p} = -\omega_\mathrm{m}$, with a repetition frequency set by $\omega_\mathrm{m}$. As the pump power increases, the number of detectable frequency comb teeth increases accordingly. Meanwhile, the entire set of comb teeth remains symmetrically distributed around $\omega_\mathrm{c2}$.

\par
For comparison, we also independently characterized the emergence and evolution of the Kerr optomechanical frequency comb in cavity-2. As shown by the red data points in Fig.~\ref{fig3}\textbf{b}, the minimum pump power $P_\mathrm{p-Kerr} = 11\ \mathrm{dBm}$ occurs at the blue sideband $\Delta_\mathrm{p} \approx \omega_\mathrm{m}$ because of maximized antidamping, triggering the optomechanical instability most efficiently.
\par
Focusing on the Floquet mode of $ m = 1 $ in Fig.~\ref{fig3}\textbf{b}, the minimum generated power for the Floquet frequency comb is approximately reduced by six orders of magnitude compared to Kerr optomechanical combs. Moreover, for the $ m = 1 $ mode, the width of the local minimum is broader and shifts toward $ m = 0 $. This occurs because, at this point, the pump not only serves as the probe of the Floquet frequency comb, but also drives the mechanical oscillator, acting similarly to the mechanism in Kerr optomechanical combs. The interaction between the pump signal and the mechanical oscillator results in a broader local minimum and the shift of this mode.

\subsection*{Effect of modulation source and coherence}
\begin{figure*}
\centering
\includegraphics[width=1\linewidth]{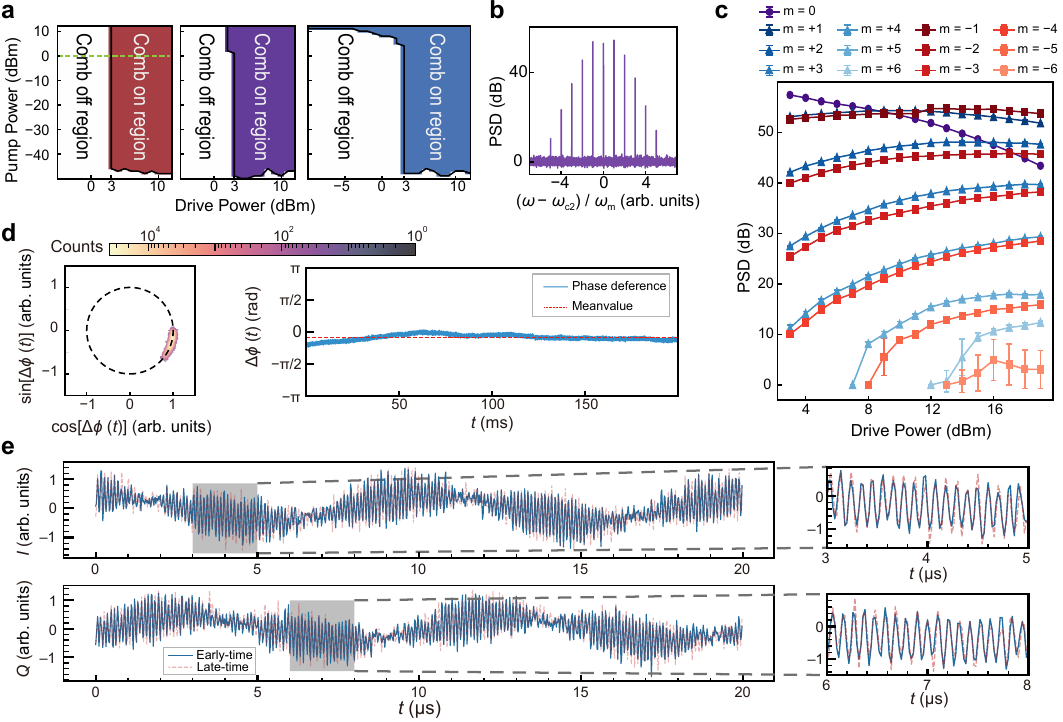}
\caption{\label{fig4}
\textbf{Parameter dependence and phase coherence.}
\textbf{a} Effect of drive and pump power on Floquet cavity frequency comb generation at $ m = -1, 0, +1 $ (Red, purple, blue). The colored regions indicate that the Floquet comb is generated, whereas the white areas represent the absence of a detectable frequency comb. The green dashed line represents the power range selected for subsequent analysis.
\textbf{b} Example frequency comb under $\Delta_\text{p}=-\omega_\text{m}$ in the frequency domain.
\textbf{c} Effect of modulation strength on the relative magnitudes of the frequency comb teeth, with error bars representing standard deviations. The gradient-colored points represent the peak values of the corresponding $m$-th comb tooth.
\textbf{d} Phasor distribution and temporal evolution of the phase deference $\Delta\phi_{-1,1}(t)$. The amplitude is normalized to its mean value and all carrier frequencies are removed. The color scale in the phasor plot represents the counts per unit area, while the red dashed line in the temporal trace indicates the mean value of $\Delta\phi_{-1,1}(t)$.
\textbf{e} Representative time slices of the 200~ms in-phase ($I$) and quadrature ($Q$) components: an early-time segment from 10~µs to 30~µs after the onset of the recorded trace (blue) and a late-time segment from 199.97~ms to 199.99~ms of the trace (red). The insets provide enlarged views of each window.
}
\end{figure*}
We investigate the impact of modulation strength on the Floquet frequency comb by setting the pump tone at $\Delta_\mathrm{p}/\omega_\mathrm{m} = -1, 0, +1$ to match the $m$-th Floquet mode. As shown in Fig.~\ref{fig4}\textbf{a}, while varying both the drive power and pump power, for $ \Delta_\mathrm{p} = -\omega_\mathrm{m} $, the frequency comb is generated only when the Floquet modulation is activated, $ P_\mathrm{d} > P_\mathrm{dc} $. The fluctuations in the minimum power value are attributed to measurement uncertainties. Notably, for $ \Delta_\mathrm{p} = 0 $, when the pump power is sufficiently high, the frequency comb is generated before $ P_\mathrm{dc} $. This is the result of the cooperative action of the optomechanical frequency comb under the external driving of two cavities. The results for $ \Delta_\mathrm{p} = \omega_\mathrm{m} $ exhibit similar behavior. At sufficiently high pump power, cavity-2 can independently generate a frequency comb via Kerr optomechanical interactions. Thus, we observe a gradual onset rather than a sharp transition. 
\par
It should be noted that, despite the absence of a threshold with respect to the pump tone, a finite total power consumption is still required to sustain self-induced oscillations.
Due to the presence of attenuators (approximately 65~dB) between the microwave sources at room temperature and the cavity input port at 10~mK (see Supplementary Note~III.B), a typical source power of 10~dBm corresponds to approximately 3.2~nW at the cavity.
If we consider the mechanical oscillator itself, the power required to resonantly drive it to a vibrational amplitude of 1~nm is only 93~fW, suggesting that the overall power consumption can be further reduced. To minimize interference from the Kerr optomechanical frequency comb in our analysis, we focus on the pump at $ \Delta_\mathrm{p} = -\omega_\mathrm{m} $. Fig.~\ref{fig4}\textbf{b} shows an example of Floquet frequency comb with $ P_\mathrm{d} = 11~\mathrm{dBm}$ and $ P_\mathrm{p} = 0~\mathrm{dBm} $.
\par
We demonstrate the effect of modulation strength for the $ \Delta_\mathrm{p} = -\omega_\mathrm{m} $ case with $ P_\mathrm{p} = 0~\mathrm{dBm} $, shown in Fig.~\ref{fig4}\textbf{c}. As the modulation strength increases, the Floquet cavity's ability to scatter photons into higher-order sidebands strengthens, resulting in a decrease in the intensity of the peak at $ m = 0 $ and an increase in the intensities of peaks at $ m = \pm 1, \pm 2, \dots $. With further increase in modulation strength, the intensity of the lower-order sidebands stabilizes and begins to decrease, while the effect of higher-order sidebands becomes more pronounced. 
At higher $P_\mathrm{d}$ values, the intensity of the $m=-1$ mode exhibits an anomalous increase due to its spectral overlap with the pump tone. We also observe an asymmetry between the two sets of sidebands, which is an intrinsic feature of the Floquet cavity frequency comb resulting from the finite pump detuning ($\Delta_\mathrm{p} \neq 0$).

We perform a preliminary verification of the coherence by measuring the phase difference 
$\Delta \phi_{-1,1}(t)$ obtained from the beat note between the $m=-1$ and $m=+1$ comb teeth. 
Fig.~\ref{fig4}\textbf{d} shows both the phasor distributions and the temporal evolution of 
$\Delta \phi_{-1,1}(t)$ under $\omega_\mathrm{p} = \omega_\mathrm{c2}$.
The phasors are tightly clustered and remain confined within a narrow angular sector, rather than being uniformly distributed over the unit circle, indicating that the phase difference between the comb lines is stable.
The temporal evolution further demonstrates bounded phase fluctuations without phase diffusion over the full $2\pi$ range.
Meanwhile, we recorded the temporal phase evolution of the $m=+1$ and $m=+2$ comb teeth and compared the segments from 10~µs to 30~µs after the start and from 30~µs to 10~µs before the end of the trace, as shown in Fig.~\ref{fig4}\textbf{e}. The phase traces at different times overlap, indicating the phase coherence and stability of the comb teeth. Additional verification of coherence was obtained through measurements of phase noise and cross-correlation between individual comb teeth (Supplementary Note~V).
\par

\section*{Discussion}

We have demonstrated that cavity Floquet engineering is a viable option for the generation of a frequency comb. While our current demonstration shows approximately ten comb teeth, this number is not an intrinsic physical limitation of the Floquet mechanism. In the experiment, the drive power was deliberately limited to avoid saturation of the cryogenic and room-temperature microwave amplifiers, which would otherwise introduce distortions and restrict the measurable comb bandwidth. The straightforward approach to increasing the number of comb teeth is to enhance the modulation strength. In our system, this can be achieved either by increasing the single-photon coupling strength to the Floquet cavity ($g_{0,2}$) or by strengthening the external drive applied to the oscillator. Apart from optomechanical driving, alternative excitation methods include capacitive excitation~\cite{reed2017Faithful, bothner2020Cavity}, piezoelectric excitation~\cite{covaci2020Piezoelectric}, and dielectric force excitation~\cite{unterreithmeier2009Universal}, which can further increase the amplitude of the oscillator.

The Floquet cavity frequency comb demonstrated here exhibits a repetition frequency of 9.1~MHz, which matches $\omega_\text{m}$. Integrating multiple mechanical oscillators on the same chip or utilizing higher-order vibration modes of the mechanical oscillators can provide additional degrees of freedom for tuning the repetition frequency of the comb. More generally, through appropriate mechanical design, the vibration frequency of mechanical oscillators can span a wide range from the kilohertz to the gigahertz regime~\cite{aspelmeyer2014Cavity}. Moreover, techniques such as injection locking~\cite{he2025coherent}, feedback control~\cite{carlson2018Ultrafast,kuang2023Nonlinear}, and the use of higher-$Q$ mechanical oscillators~\cite{liu2025Degeneracybreaking} provide promising routes further to enhance the stability of the Floquet cavity frequency comb.

Floquet cavity engineering paves the way for the efficient generation and control of frequency combs, rendering them well suited for low-power operation and compatibility with cryogenic quantum systems. For example, in superconducting quantum computing devices, components such as superconducting resonators and qubits cannot operate under high photon occupations\cite{blais2021Circuit}. Our system offers an on-chip integrated solution for low-temperature frequency conversion or frequency-division multiplexing~\cite{bock2018Highfidelity, takeuchi2024Microwavemultiplexed}. The achieved several-megahertz repetition rate is sufficient to resolve qubits, which typically exhibit kilohertz-scale linewidths~\cite{bland2025Millisecond}. Moreover, Floquet cavity frequency combs can be realized and applied in various systems exhibiting periodic modulation, such as Josephson junction~\cite{li2013Motional, bao2024Cryogenic}, trapped ions~\cite{chou2020Frequencycomb}, quantum dot~\cite{koski2018Floquet}, and molecular or atomic systems~\cite{baruch1992Ramsey, zhang2024Floquet}, particularly in the optical domain. This holds promise for extending the capabilities of frequency combs in high-dimensional quantum states~\cite{reimer2019Highdimensional}, entanglement generation~\cite{reimer2016Generation, jia2025Continuousvariable}, and integrated multi-frequency signal sources across more systems~\cite{bao2024Cryogenic, chou2020Frequencycomb}.

\par
\section*{Methods}
\subsection*{Floquet cavity}
Considering the Floquet Cavity Hamiltonian, the general form of the steady-state solution $\left|\psi_{\alpha}, t\right\rangle$ to the Schrödinger equation can be written as

\begin{equation}\label{FQ_phi}
 \left|\psi_{\alpha}, t\right\rangle=e^{-i \varepsilon_{\alpha} t}\left|u_{\alpha}, t\right\rangle,\left|u_{\alpha}, t\right\rangle=\left|u_{\alpha}, t+T\right\rangle,
\end{equation}

Here, $\varepsilon_{\alpha}$ is the quasi-energy. So the Hamiltonian can be represented as a diagonal matrix $H(t)=\sum_{n} h_{n}(t)|n\rangle\langle n|$ under the Fock basis $\{|n\rangle\}$, with matrix elements of $h_{n}(t)= n\left(\omega_{\mathrm{0}}+A \cos \Omega t\right)$. Thus, the Schrödinger equation can be simplified as

\begin{equation}\label{FQ4}
i \partial_{t} \psi_{n}(t)=n\left(\omega_{0}+A \cos \Omega t\right) \psi_{n}(t),
\end{equation}
here $\psi_{n}(t)=\langle n|\psi,t\rangle$. Combined with Equation~\eqref{FQ_phi}, the steady-state solution can be expressed as

\begin{equation}
\begin{aligned}
\left|\psi_{n}, t\right\rangle & =e^{-i n \omega_{0} t}\left|u_{n}, t\right\rangle, \\
\left|u_{n}, t\right\rangle & =e^{-i n \frac{A}{\Omega} \sin \Omega t}|n\rangle.
\end{aligned}
\end{equation}

We define $\left|u_{n m}, t\right\rangle = \left|u_{n}, t\right\rangle e^{i m \Omega t} =u_{n m}(t)|n\rangle$, with $u_{n m}(t) =e^{-i n \frac{A}{\Omega} \sin \Omega t} e^{i m \Omega t}$, which satisfies

\begin{equation}\label{FQHFSc}
 H_{F}\left|u_{n m}, t\right\rangle=\varepsilon_{n m}\left|u_{n m}, t\right\rangle.
\end{equation}

At this point, we have identified Brillouin regions over the frequency domain, where the single Brillouin region has a width of $\Omega$, and the different Brillouin regions are represented by the index $m$, called the Floquet index. And where $H_{F}=H(t)-i \partial_{t}$ is called the Floquet Hamiltonian, $\left|u_{nm}, t\right\rangle$ is the time-periodic Floquet state, and quasi-energy $\varepsilon_{n m}=n \omega_{0}+m \Omega$. The $\left|u_{n}, t\right\rangle$ can also be written as $\left|u_{n 0}, t\right\rangle$.
Under the complete basis vector set $\left.\left\{\left|n_{m n}\right\rangle\right\rangle\right\}$ with $\left|n_{m}\right\rangle=|n\rangle e^{i m \Omega t}$, the matrix elements of the Floquet Hamiltonian are

\begin{equation}\label{FQelement}
 \left.\left\langle\left\langle n_{m}\right| H_{F} \mid n_{m^{\prime}}^{\prime}\right\rangle\right\rangle=\langle n| H^{\left[m-m^{\prime}\right]}\left|n^{\prime}\right\rangle+m^{\prime} \Omega \delta_{m m^{\prime}} \delta_{n n^{\prime}},
\end{equation}
where $H^{[l]}$ is the Fourier component of $H(t)$. Since

\begin{equation}
\begin{aligned}
H(t) & =\sum_{n} n\left(\omega_{0}+A \cos \Omega t\right)|n\rangle\langle n| \\
& =\sum_{l} H^{[l]} e^{i l \Omega t},
\end{aligned}
\end{equation}
where $H^{[0]}=\sum_{n} n \omega_{c}|n\rangle\langle n|$, $H^{[\pm 1]}=\sum_{n} n \frac{A}{2}|n\rangle\langle n|$, and $H^{[l]}=0$ for $|l|>1$. Thus, the Equation~\eqref{FQelement} can be further expressed as

\begin{equation}
\begin{aligned}
\left.\left\langle\left\langle n_{m}\right| H_{F} \mid n_{m^{\prime}}^{\prime}\right\rangle\right\rangle&=\delta_{n n^{\prime}}\left(n \omega_{0} \delta_{m m^{\prime}}+n \frac{A}{2} \delta_{m, m^{\prime} \pm 1}\right)\\
&+m^{\prime} \Omega \delta_{m m^{\prime}} \delta_{n n^{\prime}},
\end{aligned}
\end{equation}

Obviously, $H_{F}$ is a block diagonal form about $n$, and we write the projection of $H_{F}$ in subspace $\left.\left\{\left|n_{m}\right\rangle\right\rangle\right\}$ with fixed $n$ as $H_{F}^{(n)}$:

\begin{equation}
H_{F}^{(n)}=\left[\begin{array}{ccccc}
\ddots & \vdots & \vdots & \vdots &\iddots \\
\cdots & n \omega_{0}-\Omega & n \frac{A}{2} & 0 & \cdots \\
\cdots & n \frac{A}{2} & n \omega_{0} & n \frac{A}{2} & \cdots \\
\cdots & 0 & n \frac{A}{2} & n \omega_{0}+\Omega & \cdots \\
\iddots & \vdots & \vdots & \vdots & \ddots
\end{array}\right] \text {, }
\end{equation}
which satisfies $H_{F}^{(n)}\left|u_{n m}\right\rangle\rangle=\varepsilon_{n m}\left|u_{n m}\right\rangle\rangle$ according to the Equation~\eqref{FQHFSc}. It can be observed that each Fock state energy level splits into $ 2m+1 $ sublevels, consistent with the periodic nature of the Floquet Hamiltonian. The off-diagonal elements, proportional to $ A/2 $, characterize the coupling strength between adjacent Floquet states, thereby determining the transition rates between different energy levels. This coupling reflects the influence of the periodic driving field and results in transitions mediated by the exchange of discrete energy quanta of size $ \Omega $. For more detailed derivations, please refer to the Supplementary Note~I.B.

\subsection*{Self-induced oscillation by blue-detuned driving}

In the blue-detuned regime, the optomechanical damping rate $ \gamma_\mathrm{opt} $ is negative. Initially, this results in an increase in the oscillator’s effective temperature. As the overall damping rate $ \gamma_\mathrm{m} + \gamma_\mathrm{opt} $ becomes negative, any small initial fluctuation will grow exponentially over time, eventually reaching a steady-state regime. These are referred to as self-induced optomechanical oscillations, where the mechanical oscillation amplitude stabilizes at a fixed value $ X_\mathrm{m} $. The governing dynamical equations are given by

\begin{equation}
 \dot{a}_1 (t) = \left[-\frac{\kappa_1}{2} + i \left( \Delta_\mathrm{d}- \frac{g_\mathrm{01}}{x_\mathrm{zpf}}x \right) \right] a_\mathrm{1}(t) + \sqrt{\kappa_\mathrm{e1}} a_{\mathrm{d}},
\end{equation}
\begin{equation}
\frac{\mathrm{d}^{2} x(t)}{\mathrm{d} t^{2}}+\gamma_{\mathrm{m}} \frac{\mathrm{~d} x(t)}{\mathrm{d} t}+\omega_{\mathrm{m}}^{2} x(t)=\frac{F(t)}{m_\mathrm{eff}},
\end{equation}
where the radiation-pressure force $F(t) =\frac{\hbar g_\mathrm{01}}{ x_{\mathrm{zpf}}}|a_\mathrm{1}(t)|^{2}$, $\kappa_1$ is the total decay rate of the cavity-1, $a_\mathrm{d}$ represents the drive photons, and $m_\mathrm{eff}$ is the effective mass of the mechanical oscillator. 

Ignoring in the initial phase $ \phi_\mathrm{m} $, we have $ x(t) = X_\mathrm{m} \cos(\omega_\mathrm{m} t) + X_\mathrm{0} $. In the steady state, we define an amplitude-dependent effective optomechanical damping rate as

\begin{equation}
 \gamma_\mathrm{opt} = -\frac{\left\langle F\dot{x} \right\rangle}{m_\mathrm{eff} \left\langle \dot{x}^2 \right\rangle},
\end{equation}
which represents the ratio of the time-averaged power input due to this force to the corresponding mechanical energy lost. The time-averaged radiation-pressure force determines the oscillation offset $ X_\mathrm{0} $, and by combining the balance condition of the overall damping rate, we obtain
\par
\begin{equation}\label{eqn-13}
 \left\{
 \begin{aligned}
 m_\mathrm{eff}\omega ^2_\mathrm{m}X_\mathrm{0} & = \left\langle F \right\rangle,\\
 \gamma _\mathrm{m} + \gamma _\mathrm{opt} & = 0.
 \end{aligned}
 \right.
\end{equation}
\par
With the ansatz, the solution of $a_\mathrm{1}(t)$ in blue-sideband drive can be written in a Fourier series $a_\mathrm{1}(t) = e^{-i \varphi(t)}\sum_{k}a_k e^{ik\omega_\mathrm{m} t}$, with coefficients:
\par
\begin{equation}
a_k = \sqrt{\kappa_\mathrm{e1}} a_{\text{d}} \frac{J_k(g_\mathrm{01} X_\mathrm{m}/x_\mathrm{zpf}\omega_\mathrm{m})}{i k\omega_\mathrm{m} + \frac{\kappa}{2} - i(\omega_\mathrm{m}- g_\mathrm{01} X_\mathrm{0}/x_\mathrm{zpf})},
\end{equation}
and the global phase $\varphi(t) = (g_\mathrm{01} X_\mathrm{m}/x_\mathrm{zpf} \omega_\mathrm{m}) \sin(\omega_\mathrm{m} t)$. By substituting this Fourier series into Equation~\eqref{eqn-13}, we obtain the implicit equation for $ X_\mathrm{0} $ and $ X_\mathrm{m} $:

\begin{equation}
 \left\{
 \begin{aligned}
 X_\mathrm{0} & = \frac{\hbar g_\mathrm{01}}{ x_{\mathrm{zpf}}m_\mathrm{eff}\omega ^2_\mathrm{m}}\sum_k|a_k|^2,\\
 X_\mathrm{m} & = \frac{2\hbar g_\mathrm{01}}{x_\mathrm{zpf} m_\mathrm{eff} \omega_\mathrm{m} \gamma_\mathrm{m}}\mathbf{Im}{\sum_k a_k^* a_{k+1}}.
 \end{aligned}
 \right.
\end{equation}

The series can be efficiently summed numerically to obtain the explicit dependence of $ X_\mathrm{m} $ and $ X_\mathrm{0} $ on the system parameters. Mathematically, the onset of small-amplitude oscillations, starting from $ A = 0 $, represents a Hopf bifurcation. In this regime, $ A \propto \sqrt{|a_\mathrm{d}|^2 - |a_\mathrm{d}|_\mathrm{th}^2} $, where $ |a_\mathrm{d}|_\mathrm{th}^2 $ is the threshold value~\cite{marquardt2006Dynamical, ludwig2008Optomechanical, miri2018Optomechanical, krause2015Nonlinear}.
For more detailed derivations, please refer to the Supplementary Note~IV.\@

\section*{Data availability}
The data supporting the findings of this study are available within the article and its Supplementary Information files. Data are available from the corresponding authors upon request. Source data are provided with this paper.

\section*{Acknowledgments}
The authors sincerely thank fruitful discussions with Kaiye~Shi and Qiongyi~He on the Floquet Hamiltonian.
\\
\section*{Funding}
Y.L. discloses support for the research of this work from Beijing Natural Science Foundation (No.~Z240007), National Natural Science Foundation of China (No.~12374325 and No.~92365210), CAST Young Elite Scientists Sponsorship Program (No.~2023QNRC001), and Beijing Municipal Science and Technology Commission (No.~Z221100002722011).
J.Z. discloses support for publication of this work from the Innovation Program for Quantum Science and Technology (No.~2021ZD0302200),
the Chinese Academy of Sciences (No.~GJJSTD20200001),
the National Key R\&D Program of China (No.~2021YFB3202800).
M.A.S. discloses support for publication of this work from European Research Council (contract 101019712), Research Council of Finland (Finnish Quantum Flagship, No.~358877).
S.W., C.W., M.H.J.d.J., and L.M.d.L. declare no relevant funding.
\\
\section*{Author contributions}
S.W. performed the experiments and carried out the theoretical analysis. C.W. fabricated the devices. M.H.J.d.J. and S.W. developed the numerical simulations. Y.L. conceived the idea, coordinated the project, and supervised the research. S.W., C.W., M.H.J.d.J., L.M.L., J.Z., M.A.S. and Y.L. contributed to the analysis of the measurement results, wrote the manuscript, and provided their final approval for publication.
\section*{Competing interests}
The authors declare no competing interests.
\end{document}